\documentclass{article}
\usepackage{graphicx}
\usepackage{amsmath}
\usepackage{amsfonts}
\usepackage{amssymb}
\newtheorem{theorem}{Theorem}

\newtheorem{definition}[theorem]{Definition}

\newtheorem{proposition}[theorem]{Proposition}

\begin{document}

\title{New Constructions in Local Quantum Physics\footnote{Talk presented at the
symposium ``Rigorous quantum field theory'' in honor of J. Bros, Paris, July
2004, submitted for publication in the symposium proceedings.}}
\author{Bert Schroer\\CBPF, Rua Dr. Xavier Sigaud 150\\222-180 Rio de Janeiro, Brazil\\and Institut fuer Theoretische Physik der FU-Berlin}
\date{January 4, 2005}
\maketitle
\begin{abstract}
Among several ideas which arose as consequences of modular localization there
are two proposals which promise to be important for the classification and
construction of QFTs. One is based on the observation that wedge-localized
algebras under certain conditions have particle-like generators with simple
properties and the second one uses the structural simplification of
wedge-localized algebras in the holographic lightfront projection.
Factorizable d=1+1 models permit to analyse the interplay between
particle-like aspects and chiral field properties of lightfront holography.

Pacs 11.10.-z, 11.55.-m
\end{abstract}

\section{How modular theory entered particle physics}

The following introductory remarks about the history and the physical content
of modular theory are intended to be helpful to understand the recent role of
modular localization in the classification and construction of models of QFT
without the use of Lagrangian quantization.

\subsection{Remarks about history of modular localization}

The beginnings of modular theory date back to the second half of the 1960s
when two independent ideas, one from mathematics and one from particle
physics, merged together \cite{Bo1}. On the mathematical side the Japanese
mathematician Tomita generalized a concept, which before was only studied in
the special context of the Haar measure (''unimodular'') in group algebra
theory, to the general setting of von Neumann algebras. At the same time three
physicists, Haag, Hugenholtz and Winnink \cite{HHW}, found a conceptual
framework for the direct field theoretic description of the thermodynamic
limit (``open systems'') in terms of operator algebras and their commutants
\cite{Haag}. Their important contribution, which became immediately
incorporated by Takesaki into Tomita's modular theory, was the realization
that the KMS condition (introduced by Kubo, Martin and Schwinger as a
computational tool) acquired a fundamental conceptual significance in their
new thermal setting.

It took another decade in order to appreciate the geometric significance of
this modular formalism for the problem of localization of algebras and states
in QFT \cite{B-W}. This was preceded by an important mathematical application
in the classification of type III von Neumann algebras by A. Connes
\cite{Connes} and followed up by a theorem of Driessler \cite{Driessler}
stating that wedge-localized algebras are factors of type III$_{1}.$ As a
consequence double cone algebras in conformal invariant theories inherit this
property\footnote{In conformal theories double cone algebras are conformally
equivalent to wedge algebras and therefore inherit the hyperfinite typ
III$_{1}$ property.
\par
{\small {}}}. Later refinements supported the idea that compactly localized
subalgebras in QFT are isomorphic to the unique hyperfinite type III$_{1}$
factor. For more detailed reviews of modular theory from the mathematical
physics viewpoint we refer to \cite{Bo1}\cite{Su}\cite{Yng}

Although hyperfinite type III$_{1}$ algebras appear at first sight (as a
result of this uniqueness) in a certain sense as void of structure as points
in geometry, they are in other aspects much richer since they contain
subalgebras of all types and one can form nontrivial intersections from copies
placed into different positions within a common Hilbert space $H$. In fact we
know from later developments of algebraic QFT that the full richness of a
model of QFT is encoded in the notion of a net of spacetime-indexed von
Neumann algebras as subalgebras of $B(H)$ \cite{Haag}.

The net result of this thread of ideas, which culminated in the mathematical
identification of simple building blocks of QFT, is interesting from many
viewpoints. From a philosophical standpoint it tells us that the algebraic
aspects of QFT comply perfectly with Leibniz's dictum that reality emerges
from the relation between indecomposable entities (``monades'') and not from
their individual position with respect to an absolute outside reference.

This is not the first time that philosophical ideas of Leibniz became relevant
in physics. In Einstein's ``hole argument'' \cite{Norton} it played a
significant role in the birth of general relativity; in particular it helped
Einstein (and independently Hilbert) to overcome a misconception about how the
local covariance of the Einstein-Hilbert field equations and the Newtonian
limit fit together. By upholding the local covariance principle, i.e. the idea
that local isomorphism classes of isometric diffeomorphisms replace the global
notion of an absolute Minkowski spacetime inertial frame of special
relativity, Einstein realized that his difficulties to obtain agreement with
the Newtonian limit came from a computational misconception.

In fact it was shown recently that the Leibniz viewpoint of physical reality
emerging from relations between entities rather than from positions in a
pre-assigned absolute ``inertia ether'' can actually be extended in order to
combine the quantum algebraic modular aspects with the classical covariance
principle into a ``local (quantum) covariance principle'' \cite{B-F-V}%
\cite{review} which achieves background independence on the algebraic level
whereas states generically change under local diffeomorphisms but in such a
way that the affiliation with a folium is maintained. This places QFT in
curved spacetime much closer to a still elusive background-independent quantum
gravity than ever before.

In the following we will argue that the ``monades'' of QFT are the
wedge-localized algebras which (thanks to Driessler's work \cite{Driessler})
are known to be isomorphic copies of hyperfinite type III$_{1}$factor. In
order to avoid lengthy terminology we will refer to the basic hyperfinite type
III$_{1}$von Neumann algebra as the monade algebra (MA). The most convincing
affirmation of this way of viewing QFT as arising from wedge localized MA is
the fact that models of quantum field theory can be completely specified by
positioning a finite number of operator algebra copies of the MA into suitably
chosen relative positions within a common Hilbert space\footnote{We will use
the terminology MA also for the positioned operator algebra copies of the
basic MA.} \cite{K-W}.

This way of looking at QFT permits a particularly natural intrinsic
formulation in low dimensional QFT. In the case of d=1+1 the ``modular
inclusion'' of two MA specifies all data needed to characterize a specific QFT
in terms of the structure of its Poincar\'{e} covariant nets and for d=1+2 one
achieves a complete characterization in terms of ``modular intersection'' of
three copies of the MA. Modular positions which are associated with a
characterization of higher dimensional QFT models are also known \cite{K-W},
but in their present formulation they appear less natural i.e. more concocted
in order to generate the desired Poincar\'{e} symmetry structure of Minkowski spacetime.

Accepting the philosophical, conceptual and mathematical implications of such
a viewpoint, one may ask whether this approach guided by Leibniz's philosophy
is just an esoteric new way of looking at particle physics or if it also has
constructive clout, i.e. whether one can actually classify models of QFT and
elaborate a realistic scenario of their construction along those lines.
Admittedly the apparent simplicity of generating QFT from the positioning of a
finite number of MAs is somewhat deceiving; the problem of an intrinsically
formulated relative positioning of MAs is actually quite hard since the
appropriate concepts and mathematical tools are to a large extend still
missing. Already the characterization of one MA in Hilbert space i.e. in the
setting of local quantum physics of massive particles the description
$\mathcal{A}(W)\subset_{P}B(H)$ is a difficult problem; here $\mathcal{A}(W)$
denotes a wedge-localized MA for a fixed wedge, $B(H)$ is the algebra of
bounded operators on Fock space of massive particles obtained by scattering
theory (assuming asymptotic completeness) and the subscript $P$ indicates that
the inclusion is meant in the extended sense that also the action of the
Poincar\'{e} group on it (which creates a family of wedge-localized MAs for
all Ws) is known.

There are two situations in which this positioning of the MA is reasonably
simple and the construction of the net (and its generating pointlike field
coordinatizations) can actually be carried out. These are the interaction-free
theories whose one particle components are described in terms of Wigner
representations of elementary systems and d=1+1 factorizing models. For
general interacting theories the idea of lightfront holography turns out to be
very helpful because it suggests to classify and construct wedge algebras in
terms of their lightfront holographies. These problems will be addressed in
this paper.

In the remainder of this introduction the modular approach to the
interaction-free QFTs will be given; as a result of its simplicity it serves
well as a pedagogical introduction into the setting of modular localization.

\subsection{Modular construction of interaction-free QFT}

This construction via modular localization proceeds in three steps as follows
\cite{B-G-L} \cite{Mund}\cite{F-S}\cite{MSY2}

\begin{enumerate}
\item  Fix a reference wedge region, e.g. $W_{R}=\left\{  x\in\mathbb{R}%
^{4};x^{1}>\left|  x^{0}\right|  \right\}  $ and use the Wigner positive
energy representation of the $W_{R}$-affiliated boost group $\Lambda_{W_{R}%
}(\chi)$ and the $x^{0}-x^{1}-$reflection $j_{W}$\footnote{The reflection on
the edge of the wedge is related to the total TCP reflection by a $\pi
$-rotation around the $x_{1}$-axis. } along the edge of the wedge $j_{W_{R}}$
in order to define the following antilinear unbounded closable operator (with
$closS=clos\Delta^{\frac{1}{2}}$). Retaining the same notation for the closed
operators, one defines
\begin{align}
S_{W_{R}} &  :=J_{W_{R}}\Delta^{\frac{1}{2}}\\
J_{W_{R}} &  :=U(j_{W_{R}}),\;\Delta^{it}:=U(\Lambda_{W_{R}}(2\pi t))\nonumber
\end{align}
The commutativity of $J_{W_{R}}$ with $\Delta^{it}$ together with the
antiunitarity of $J_{W_{R}}$ yield the property which characterize a Tomita
operator\footnote{Operators with this property are the corner stones of the
Tomita-Takesaki modular theory \cite{Takesaki} of operator algebras. Here they
arise in the spatial Rieffel-van Daele spatial setting \cite{Rieffel} of
modular theory from a realization of the geometric Bisognano-Wichmann
situation within the Wigner representation theory.} $S_{W_{R}}^{2}\subset1$
whose domain is identical to its range$.$ Such operators are completely
characterized in terms of their $+1$ real eigenspaces which in the present
context amounts to real standard subspace $K(W_{R})$ of the Wigner
representation space $H$
\begin{align}
K(W_{R}) &  :=\left\{  \psi\in H,\,S_{W_{R}}\psi=\psi\right\}  \label{real}\\
\overline{K(W_{R})+iK(W_{R})} &  =H,\,\,K(W_{R})\cap iK(W_{R})=0\nonumber\\
J_{R}K(W_{R}) &  =K(W_{R})^{\bot}=:K(W_{R})^{\prime}\nonumber
\end{align}
$K(W_{R})$ is closed in $H$ whereas the complex subspace spanned together with
the -1 eigenspace $iK(W_{R})$ is the dense domain of the Tomita operator
$S_{W_{R}}$ and forms a Hilbert space in the graph norm of $S_{W_{R}}$. The
denseness in $H$ of this span $K(W_{R})+iK(W_{R})$ and the absence of
nontrivial vectors in the intersection $K(W_{R})\cap iK(W_{R})$ is called
``standardness''. The right hand side in the third line refers to the
symplectic complement i.e. a kind of ``orthogonality'' in the sense of the
symplectic form $Im(\cdot,\cdot).$ The application of Poincar\'{e}
transformations to the reference situation generates a consistent family of
wedge spaces $K(W)=U(\Lambda,a)K(W_{R})$ if $W=(\Lambda,a)W_{R}.$ These
subspaces carry a surprising amount of informations about local quantum
physics; their structure even preempts the spin-statistics connection by
producing a mismatch between the symplectic and the geometric complement
($W^{\prime}$ denotes the causal complement in terms of Minkowski space
geometry) which is related to the spin-statistics factor \cite{Mund}%
\cite{F-S}
\begin{align}
K(W)^{\prime} &  =ZK(W^{\prime})\\
Z^{2} &  =e^{2\pi is}\nonumber
\end{align}
Another surprising fact is that the modular setting prepares the ground for
the field theoretic on-shell crossing property, since the equation
characterizing the real modular localization subspaces for general spin reads
\begin{equation}
\left(  S_{W_{R}}\psi\right)  \left(  p\right)  =\left(  J\Delta^{\frac{1}{2}%
}\psi\right)  (p)=V(p)\overline{\psi(-p)}=\psi(p)
\end{equation}
i.e. the complex conjugate of the analytically continued wave function (but
now referring to the charge-conjugate situation$)$ is up to a p-dependent
matrix $V(p)$ which acts on the spin indices (see \cite{Mund} formula (2.14))
equal to the original wave function. One easily checks that this unbounded
antiunitary operator S acts Lorentz-invariantly on a certain domain
\cite{Munds}.

\item  The sharpening of localization is obtained by intersecting wedges in
order to obtain real subspaces as causally closed subwedge
regions\footnote{Instead of working with real subspaces one can formulate the
content of the spatial modular theory by only using the Tomita operators S and
their domains \cite{Mund2}.}:
\begin{equation}
K(\mathcal{O}):=\cap_{W\supset\mathcal{O}}K(W)\label{int}%
\end{equation}
The crucial question is whether they are ``standard''. According to an
important theorem of Brunetti, Guido and Longo \cite{B-G-L} standardness
universally holds for spacelike cones $\mathcal{O}=\mathcal{C}$ \ in all
positive energy representations. In case of finite spin/helicity
representations the standardness also holds for intersections leading to
(arbitrary small) double cones $\mathcal{D}$. \ In those cases where the
double cone localized spaces with pointlike ``cores'' are trivial (massless
infinite spin \cite{M-S-Y}, massive d=1+2 \ anyons \cite{Mund2}), the smallest
localization regions are spacelike cones with semiinfinite strings as cores.
Without loss of generality one may restrict localization regions to convex
causally complete regions.

\item  In the absence of interactions the transition from free particles to
localized operator algebras is most appropriately done in a functorial way by
applying the Weyl (CCR) (or in case of halfinteger spin the CAR functor) to
the localization K-spaces\footnote{To maintain simplicity we limit our
presentation to the bosonic situation and refer to \cite{Mund}\cite{F-S} for
the general treatment.}:
\begin{align}
\mathcal{A(O)} &  :=alg\left\{  Weyl(\psi)|\,\,\psi\in K(\mathcal{O})\right\}
\label{alg}\\
Weyl(f) &  :=expi\left\{  a^{\ast}(\psi)+h.a.\right\}  \nonumber
\end{align}
where $a^{\#}(\psi)$ are the creation/annihilation operators of particles in
the Wigner wave function $\psi.$ The functorial relation between real
subspaces and von Neumann algebras preserves the causal localization structure
\cite{L-R-T} and commutes with the process of improvement of localization
through the formation of intersections.
\end{enumerate}

For later purposes we introduce the following definition \cite{S2}.

\begin{definition}
A vacuum-polarization-free generator (PFG) for a region $O$ is an operator
affiliated with the algebra $\mathcal{A}$($\mathcal{O}$) which created a
vacuum-polarization-free one-particle vector
\begin{align}
&  G\,\eta\text{ }\mathcal{A(O)}\\
G\Omega &  =1-particle\nonumber
\end{align}
\end{definition}

Since these wedge algebra-affiliated operators $G$\ are generally unbounded,
one has to comment on their domain properties. We will assume that they admit
a dense domains which, similar to smeared Wightman fields \cite{S-W}, is
stable under translations. This definition permits to characterize the
presence of interactions by the interaction induced vacuum polarization as a
result of the following statement

\begin{proposition}
The existence of subwedge-localized PFGs characterizes interaction-free theories.
\end{proposition}

The proof uses the fact that PFGs are on-shell (weak solutions of the
Klein-Gordon equation \cite{B-B-S}) as well subwedge-localized; the analytic
argument is analogous to that of the theorem about the equality of a two-point
function with that of a free field implying the equality of the associated
covariant field with a free field \cite{S-W} (the restriction to covariance
and pointlike localization is easily seen to be not necessary \cite{Stein}%
\cite{Mund1}). The existence of wedge-localized PFGs $G\eta\mathcal{A}(W)$ is
a consequence of modular theory, but their domain $dom(G)$ is generally not
stable under all translations (only under those translations which transform
the wedge into itself). Such PFGs do not admit a Fourier transform i.e. they
are not tempered \cite{B-B-S}. Hence in the presence of interactions the
particle localization through the application of localized operators to the
vacuum is weakened; according to the previous proposition the QFT cannot
localize particles in subwedge regions. Accordingly the functorial relation
between particle and field localization breaks down and one has to look for a substitute.

In the next section we will show that the requirement that wedge-localized $G$
\ fulfill the domain properties of the definition (i.e. are ``tempered'')
leads to an explicit characterization of the associated wedge-localized
algebras in terms of a simple algebraic structure of their generators. This
amounts to the complete knowledge of the QFT in the sense of its algebraic
net. Namely it can be shown that the knowledge of generators of wedge algebras
together with the knowledge how Poincar\'{e} transformations acts on this
reference wedge algebra and generate the family of all wedge algebras in
different spacetime position is sufficient to build up the complete net of
algebras through the formation of intersections of wedge algebras (in analogy
to (\ref{int})). The existence of tempered wedge-localized PFGs is quite
restrictive and only admits models in d=1+1 with a purely elastic scattering
matrix \cite{B-B-S}. Examples are obtained by Fourier transforming generating
operators of Zamolodchikov-Faddeev algebras (\ref{PFG}) and there are reasons
to believe that the d+1+1 factorizing models exhaust the possibilities for
tempered PFGs. Knowing the PFG generators explicitly as one does in these
models, one can construct the net and its local field generators (which are of
course much more involved than the non-local wedge generators).

In the third section the idea of lightfront holographic projection will be
used in order to classify wedge algebras in terms of extended chiral algebras.
The problem of classifying such algebras seem to be not much more difficult
than that for usual chiral algebras.The unsolved problems of inverse
lightfront holography i.e. the problem of reconstructing ambient algebras from
their holographic projections is the main obstacle in the general
classification\&construction. Here again the restriction to d=1+1 factorizing
models turns out to be very helpful.

\section{Modular localization and the bootstrap-formfactor program}

The various past attempts at S-matrix theories which aimed at direct
constructions of scattering data without the intermediate use of local fields
and local observables provide illustrations of what is meant by an
``on-shell'' approach to particle physics. The motivation behind such
proposals was first spelled out by Heisenberg \cite{Heisenberg}. It consisted
in the hope that by limiting oneself to particles and their mass-shells, one
avoids (integration over) fluctuations on a scale of arbitrarily small
spacelike distances causing ultraviolet divergencies whose appearance at the
pre-renormalization days of Heisenberg's S-matrix proposal were seen as an
incurable disease of QFT. The main purpose of staying close to particles and
using scattering concepts (``on-shell'') is the avoidance of inherently
singular objects as pointlike fields in calculational steps. This is certainly
a reasonable aim independent of whether one believes or not that a formulation
of interactions in terms of singular pointlike fields exists for d=1+3 QFT in
the mathematical physics sense.

Since the early 1950s, in the aftermath of renormalization theory, the
relation between particles and fields received significant elucidation through
the derivation of time-dependent scattering theory. In the course of this it
also became clear that Heisenberg's S-matrix proposal had to be amended by the
addition of the crossing property i.e. a prescription of how to analytically
continue particle momenta on the complex mass shell in order to relate matrix
elements of local operators between incoming ket and outgoing bra states with
a fixed total sum of incoming and outgoing particles as different boundary
values of one analytic ``masterfunction''. In physical terms crossing allows
to relate matrix elements describing real particle creation (with particles in
both the incoming ket- and outgoing bra-states) to the vacuum polarization
matrix elements where the ket-state (or the bra state) is the vacuum vector.

Whereas Heisenberg's requirements of Poincar\'{e} invariance, unitarity and
cluster factorization on a relativistic S-matrix can also be implemented in a
``direct particle interaction'' scheme \cite{Coester}\cite{cluster}, the
implementation of crossing is conceptually related to the presence of vacuum
polarization for which QFT with its micro-causality is the natural arena.

The LSZ time-dependent scattering theory and the associated reduction
formalism relates such a matrix element (referred to as a generalized
formfactor) in a natural way to one in which an incoming particle becomes
``crossed'' into an outgoing anti-particle on the backward real mass shell; it
is at this point where analytic continuation from a positive energy physical
process enters. In this setting the S-matrix is the formfactor of the identity operator.

The important remark here is that the use of particle states requires the
restriction of the analytic continuation to the complex mass shell
(``on-shell''). It was Bros\footnote{Since the issue of crossing constitutes
an important property of the present paper, it is particularly appropriate to
dedicate this work to Jacques Bros on the occasion of his 70$^{th}$ birthday.}
in collaboration with Epstein and Glaser \cite{B-E-G} who gave the first
rigorous proofs of crossing in special configurations. In the special case of
the elastic scattering amplitude, the crossing of only one particle from the
incoming state has to be accompanied by a reverse crossing of one of the
outgoing particles in order to arrive at a physical process allowed by
energy-momentum conservation\footnote{This crossing of a pair of particles
from the in/out elastic configuration is actually the origin of the
terminology ``crossing'' and was the main object of rigorous analytic
investigations.}.

A derivation of crossing in the setting of QFT for general multi-particle
scattering configurations and for formfactors (as one needs it for the
derivation of a bootstrap-formfactor program, see later) from the general
principles of local quantum physics does not yet exist. It is not clear to me
whether the present state of art in algebraic QFT would permit to go
significantly beyond the old but still impressive results quoted before.

The crossing property became the cornerstone of the so-called bootstrap
S-matrix program and several ad hoc representations of analytic scattering
amplitudes were proposed (Mandelstam representation, Regge poles...) in order
to incorporate crossing in a more manageable form.

The algebraic basis of the bootstrap-formfactor program for the special family
of d=1+1 factorizable theories is the validity of a momentum space
Zamolodchikov-Faddeev algebra \cite{Z}. The operators of this algebra are
close to free fields in the sense that their Fourier transform is on-shell
(see (\ref{PFG}) in next section), but unlike the latter they are not local in
the pointlike sense. A closer look reveals that they are localizable in the
weaker sense\footnote{An operator which is localizable in a certain causally
closed spacetime region is automatically localized in any larger region but
not necessarily in a smaller region. The unspecific terminology ``non-local''
in the literature is used for any non pointlike localized field.} of spacetime
wedge regions \cite{S2}\cite{Lech}. In fact the existence of \ such Fourier
transformable (``tempered'') wedge-localized PFGs, which implies the absence
of real particle creation through scattering processes \cite{B-B-S}, turns out
to be the prerequisite for the success of the bootstrap-formfactor program for
factorizable models in which one uses only formfactors and avoids
(short-distance singular) correlation functions.

According to an old structural theorem which is based on certain analytic
properties of a field theoretic S-matrix \cite{Aks}\cite{B-B-S},
interaction-induces vacuum polarization without real particle creation is only
possible in d=1+1 theories. This in principle leaves the possibility of direct
3- or higher- particle elastic processes beyond two particle scattering. But
an argument by Karowski based on formfactor crossing\footnote{I am indebted to
M. Karowski for this argument.} shows that the nonvanishing of higher
connected elastic contributions would even in d=1+1 be inconsistent with the
absence of real particle creation. In this sense the Z-F algebra structure,
which is at the heart of factorizing models, turns out to be a consequence of
special properties of modular wedge-localized PFGs; a fact which places the
position of the factorizing models within general QFT into sharper focus. The
crossing property of the two-particle scattering amplitude is a consistency
prerequisite for the formfactor crossing. Providing a special illustration of
the previous general unicity argument of inverse scattering based on crossing,
the bootstrap formfactor approach associates precisely one QFT in the sense of
one local equivalence class of fields (or one net of localized operator
algebras) to a prescribed factorizing S-matrix.

In agreement with the philosophy underlying AQFT, which views pointlike fields
as coordinatizations for generators of localized algebras, the algebraic
bootstrap-formfactor construction for d=1+1 factorizing models constructs
coordinatization independent double-cone algebras by computing intersections
of wedge algebras. The nontriviality of a theory is then tantamount to the
nontriviality ($\neq C1$) of such intersections\footnote{A nontrivial
intersection could however be associated to a sub-theory. }. The computation
of a basis of pointlike field generators of these algebras is analogous to
(but more involved than) the construction of a basis of composites of \ free
fields in the form of Wick polynomials. As we saw before for noninteracting
theories, the functorial description of the algebras (\ref{alg}) based on
modular localization is conceptually simpler than the use of free fields and
their local equivalence class of Wick-ordered composites e.g. one is not
obliged to introduce a non-intrinsic Wick basis in order to parametrize the
set of all pointlike fields.

The crossing property is crucial for linking scattering data with off-shell
operators spaces. As explained in the previous section, it relates the
multi-particle component of vectors obtained by one-time application of a
localized (at least wedge-localized) operator to the vacuum with the connected
formfactors of that operator. It is important to note that in factorizing
models crossing is not an assumption but rather follows from the properties of
tempered PFGs for wedge algebras similar to crossing of formfactors for
composite operators of free fields \cite{cross2}.

In the following some of the details of wedge-localized PFGs and their
connections with the Zamolodchikov-Faddeev algebra structure are presented. In
the simplest case of a scalar chargeless particle without bound
states\footnote{A situation which in case of factorizing models with variable
coupling (e.g. the massive Thirring model) can always be obtained by choosing
a sufficiently small coupling. Bound state poles in the physical $\theta
$-strip require nontrivial changes of the algebraic formalism.{\small  }} the
wedge generators are of the form \cite{S2}%

\begin{align}
\phi(x) &  =\frac{1}{\sqrt{2\pi}}\int(e^{ip(\theta)x(\chi)}Z(\theta
)+h.c.)d\theta\label{PFG}\\
Z(\theta)Z^{\ast}(\theta^{\prime}) &  =S^{(2)}(\theta-\theta^{\prime})Z^{\ast
}(\theta^{\prime})Z(\theta)+\delta(\theta-\theta^{\prime})\nonumber\\
Z(\theta)Z(\theta^{\prime}) &  =S^{(2)}(\theta^{\prime}-\theta)Z(\theta
^{\prime})Z(\theta)\nonumber
\end{align}
Here $p(\theta)=m(ch\theta,sh\theta)$ is the rapidity parametrization of the
d=1+1 mass-shell and $x=r(sh\chi,ch\chi)$ parametrizes the right hand wedge in
Minkowski spacetime. $S^{(2)}(\theta)$ is a structure function of the Z-F
algebra which is a nonlocal $^{\ast}$-algebra generalization of canonical
creation/annihilation operators. The notation preempts the fact that
$S^{(2)}(\theta)$ is the analytic continuation of the physical two-particle
S-matrix $S^{(2)}(\left|  \theta\right|  )$ which via the factorization
formula determines the general scattering operator $S_{scat}$ (\ref{fac}). The
unitarity and crossing of $S_{scat}$ follows from the corresponding
two-particle properties which in terms of the analytic continuation are
$S^{2}(z)^{\ast}=S^{(2)}(-z)$ (unitarity) and $S^{(2)}(z)=S^{(2)}(i\pi-z)$
(crossing) \cite{KTTW}. The $Z^{\ast}(\theta)$ operator applied repeatedly to
the vacuum in the natural order $\theta_{1}>\theta_{2}>...>\theta_{n}$ is by
definition the outgoing canonical Fock space creation operators whereas the
re-ordering from any other ordering has to be calculated according to the Z-F
commutation relations e.g.
\begin{equation}
Z^{\ast}(\theta)a^{\ast}(\theta_{1})...a^{\ast}(\theta_{n})\Omega=\prod
_{i=1}^{k}S^{(2)}(\theta-\theta_{i})a^{\ast}(\theta_{1})..a^{\ast}(\theta
_{k})a^{\ast}(\theta)a^{\ast}(\theta_{k+1})..a^{\ast}(\theta_{n})\Omega
\end{equation}
where $\theta<\theta_{i}$ $i=1..k,\,\theta>\theta_{i}$ $i=k+1,..n.$ The
general Zamolodchikov-Faddeev algebra is a matrix generalization of this structure.

It is important not to identify the Fourier transform in (\ref{PFG}) of the
momentum with a localization variable. Although the $x$ in $\phi(x)$ behaves
covariantly under Poincar\'{e} transformations, it is not marking a causal
localization point; in fact it is a non-local variable in the sense of the
standard use of this terminology\footnote{The world local is reserved for
``commuting for spacelike distances''. In this work we are dealing with
non-local fields which are nevertheless localized in causally complete
subregions (wedges, double cones) of Minkowski spacetime.}. It is however
wedge-localized in the sense that the generating family of operator for the
right-hand wedge $W$ Wightman-like (polynomial) algebra $alg\left\{
\phi(f),suppf\subset W\right\}  $ commutes with the TCP transformed algebra
$alg\left\{  J\phi(g)J,suppg\subset W\right\}  $ which is the left wedge
algebra \cite{Lech}
\begin{align}
&  \left[  \phi(f),J\phi(g)J\right]  =0\label{wedge}\\
J &  =J_{0}S_{scat}\nonumber
\end{align}
Here$\ J_{0}$ is the TCP symmetry of the free field theory associated with
$a^{\#}(\theta)$ and $S_{scat}$ is the factorizing S-matrix which on
(outgoing) n-particle states has the form
\begin{equation}
S_{scat}a^{\ast}(\theta_{1})a^{\ast}(\theta_{2})...a^{\ast}(\theta_{n}%
)\Omega=\prod_{i<j}S^{(2)}(\theta_{i}-\theta_{j})a^{\ast}(\theta
_{2})...a^{\ast}(\theta_{n})\Omega\label{fac}%
\end{equation}
if we identify the $a^{\#}(\theta)$ with the incoming creation/annihilation
operators. It is then possible to give a rigorous proof \cite{Lech} that the
Weyl-like algebra generated by exponential unitaries is really wedge-localized
and fulfills the Bisognano-Wichmann property
\begin{align}
\mathcal{A}(W) &  =alg\left\{  e^{i\phi(f)}\,|\,\;suppf\subset W\right\}
\label{BW}\\
\mathcal{A}(W)^{\prime} &  =J\mathcal{A}(W)J=\mathcal{A}(W^{\prime})\nonumber
\end{align}
where the dash on operator algebras is the standard notation for the von
Neumann commutant and the dash on spacetime regions stands for the causal
complement. Within the modular setting the relative position of the causally
disjoint $A(W^{\prime})$ depends via $S_{scat}$ on the dynamics. The operator
TCP operator $J$ is the (antiunitary) ``angular'' part of the polar
decomposition of Tomita's algebraically defined unbounded antilinear
S-operator with the following characterization
\begin{align}
SA\Omega &  =A\Omega,\,\,A\in\mathcal{A}(W)\\
S &  =J\Delta^{\frac{1}{2}},\,\,\Delta^{it}=U(\Lambda(-2\pi t))\nonumber
\end{align}
with $\Lambda(\chi)$ being the Lorentz boost at the rapidity $\chi.$

At this point the setup looks like a relativistic quantum mechanics since the
$\phi(f)$ (similar to genuine free fields if applied to the vacuum) do not
generate vacuum polarization clouds. The advantage of the algebraic modular
localization setting is that interaction-caused vacuum polarization is
generated by algebraic intersections which is in agreement with the intrinsic
definition of the notion of interaction presented in terms of PFGs in the
previous section
\begin{align}
\mathcal{A}(D) &  \equiv\mathcal{A}(W)\cap\mathcal{A}(W_{a}^{\prime
})=\mathcal{A}(W)\cap\mathcal{A}(W_{a})^{\prime}\,\label{rel}\\
D &  =W\cap W_{a}^{\prime}\nonumber
\end{align}
This is the operator algebra associated with a double cone $D$ (which for
convenience is chosen symmetric around the origin by intersecting with the
causal complement of a translated wedge algebra). Note the difference from the
quantization approach, where pointlike localized fields are used from the
outset and the sharpening of localization of smeared products of fields is
simply achieved by the classical step of restricting the spacetime support of
the test functions. 

The problem of computing intersected von Neumann algebras is in general not
only difficult (since there are no known general computational techniques) but
also very unusual as compared to methods of standard quantization. There is a
well-founded hope that one can solve the existence problem of factorizing
models by showing the nontriviality of the intersections $A(D)$ \cite{Bu-Le}.

This task becomes more amenable if one considers instead of operators their
formfactors i.e. their matrix elements between incoming ket and outgoing bra
state vectors. In the spirit of the LSZ formalism one can then make an Ansatz
in form of a power series in $Z(\theta)$ and $Z^{\ast}(\theta)\equiv
Z(\theta-i\pi)$ (corresponding to the power series in the incoming free field
in LSZ theory). In a shorthand notation which combines both frequency parts we
may write
\begin{equation}
A=\sum\frac{1}{n!}\int_{C}...\int_{C}a_{n}(\theta_{1},...\theta_{n}%
):Z(\theta_{1})...Z(\theta_{n}):d\theta_{1}...d\theta_{n}\label{series}%
\end{equation}
where each integration path $C$ extends over the upper and lower part of the
rim of the $(0,-i\pi)$ strip in the complex $\theta$-plane. The
strip-analyticity of the coefficient functions $a_{n}$ expresses the
wedge-localization of $A\footnote{{\small Compact localization leads to
coefficient functions which are meromorphic outside the open strip
\cite{BFKZ}.}}.$ It is easy to see that these coefficients on the upper part
of $C$ (the annihilation part) are identical to the vacuum polarization form
factors of $A$%
\begin{equation}
\left\langle \Omega\left|  A\right|  p_{n},..p_{1}\right\rangle ^{in}%
=a_{n}(\theta_{1},...\theta_{n}),\;\ \theta_{n}>\theta_{n-1}>...>\theta_{1}%
\end{equation}
whereas the crossing of some of the particles into the left hand bra state
(see the previous section) leads to the connected part of the formfactors
\begin{equation}
^{out}\left\langle p_{1},..p_{k}\left|  A\right|  p_{n},..p_{k+1}\right\rangle
_{conn}^{in}=a_{n}(\theta_{1}+i\pi,...\theta_{k}+i\pi,\theta_{k+1}%
,..\theta_{n})
\end{equation}
Hence the crossing property of formfactors is conveniently encoded into the
notation of the operator formalism (\ref{series}) in that there is only one
analytic function $a_n$ which describes the different possibilities of placing
$\theta$ on the upper or lower rim of $C.$

The presence of bound states (poles in the physical $\theta$-strip) leads to a
weakening of the wedge localization in the sense that the wedge commutativity
(\ref{wedge}) only holds between states from the subspace generated from the
``elementary'' states linearly related to (\ref{PFG}). This requires
considerable modifications of the algebraic formalism which goes beyond the
modest aims of this paper.

The essential advantage of this algebraic formalism over the calculation of
formfactors of individual fields is expected to appear if one tries to to
solve the hard problem of the mathematical existence of the theory. Whereas
the conventional way via controlling Wightman functions and checking their
properties appears hopelessly complicated (the mathematical control of the
convergence of the formfactor series (\ref{series}) has not even been achieved
in simple models), the ``modular nuclearity property'' of wedge algebras in
d=1+1 which secures the nontriviality of the intersected algebras $A(D)$
\cite{Bu-Le} seems to be well in reach \cite{Lech2}.

\section{Constructive aspects of lightfront holography}

In the previous sections it was shown how modular theory together with
on-shell concepts can be used to analyze special wedge algebras in the
presence of interactions. The constructive use was limited to the presence of
so-called tempered PFGs which in turn restricted computable models to d=1+1
factorizing theories. In this section I will present a recent proposal which
also uses modular localization ideas but tries to analyze wedge algebras in
terms of (extended) chiral theories by means of ``algebraic lightfront
holography'' (ALH). Again we limit ourselves to some intuitively accessible
remarks mainly emphasizing analogies as well as differences with the standard
formalism of QFT; for a more detailed mathematical description we again refer
to the literature \cite{cross2}.

The following comparison with the canonical formalism turns out to be helpful.
The ETCR formalism tries to classify and construct QFTs by assuming the
validity of canonical equal time commutation relations (ETCR). The
shortcomings of that approach are well-known. Even if one ignores the fact
that the ETC structure is inconsistent with the presence of strictly
renormalizable interactions\footnote{Only superrenormalizable interactions
(finite wave function renormalization) as the polynomial scalar models in
d=1+1 have fields which restrict to equal times.}, the usefulness of the ETCR
is still limited by its insensitivity with respect to interactions. One would
prefer to start with a structure which is senses the presence of interactions
and is capable to utilize the enormous amount of knowledge and structural
richness which has been obtained in studying chiral theories by providing a
concept of rich universality classes for higher dimensional QFT (instead of
just one ETCR class).

Lightfront holography tries to address this imbalance by replacing the ETCR by
the richer structure of (extended) chiral theories on the lightfront. Its main
aim is to shift the cut between kinematics and dynamics in such a way that
what has been learned by studying low dimensional theories can be used as a
kinematical input for higher dimensional models.

The holographic projection turns out to map many different interacting ambient
theories to the same holographic image; in this respect there is a certain
similarity to the better known scale invariant short distance universality
classes which are the key to the understanding of critical phenomena. But in
contradistinction to scaling universality classes which change the theory to
an associated massless theory, holographic projections live in the same
Hilbert space as the ambient theory; in fact they just organize the spacetime
aspects of a shared algebraic structure in a radically different way.

Let us briefly recall some salient points of ALH\footnote{We add this prefix
``algebraic'' in order to distinguish the present notion of holography from
the gravitational holography of t'Hooft \cite{Hooft}. }.

ALH may be viewed as a kind of conceptually and mathematically refined
``lightcone quantization'' (or ``$p\rightarrow\infty$ frame'' description).
Whereas the latter never faced up to the question of how the lightfront
quantized fields are related to the original local fields i.e. in which sense
the new description addresses the original problems posed by the ambient
theory, the ALH is conceptually precise and mathematically rigorous on this points.

It turns out that the idea of restricting fields to the lightfront is limited
to free fields and certain superrenormalizable interacting models with finite
wave function renormalization. Theories with interaction-caused vacuum
polarization which leads to Kall\'{e}n-Lehmann spectral functions with
diverging wave function renormalization do not permit lightfront restrictions
for the same reason as they do not have equal time restrictions; e.g. for
scalar fields on has\footnote{{\small It is important to realize that LF
restriction is not a pointwise local procedure. This becomes clearer within
the setting of modular localization.}}
\begin{align}
\left\langle A(x)A(y)\right\rangle  &  =\int_{0}^{\infty}\rho(\kappa
^{2})i\Delta^{(+)}(x-y,\kappa^{2})d\kappa^{2}\\
\left\langle A(x)A(y)\right\rangle |_{LF} &  \sim\int_{0}^{\infty}\rho
(\kappa^{2})d\kappa^{2}\delta(x_{\perp}-y_{\perp})\int_{0}^{\infty}\frac
{dk}{k}e^{-ik(x_{+}-y_{+})}\nonumber
\end{align}
where in passing to the second line we used the correct rule for lightfront
restriction; this is obviously not the naive one obtained by simply
restricting the coordinates in the Kall\'{e}n-Lehmann representation. To
obtain the second line, which replaces the free field $\Delta^{+}$ function by
the transverse $\delta(x_{\perp}-y_{\perp})$ delta function times the
longitudinal chiral function in the $x_{+}$ lightray variable, one starts from
the free field representation in terms of momentum space creation/annihilation
operators. In the z-t wedge region this field may be parametrized in terms of
rapidites $\chi,\theta$ as follows:%
\begin{align}
&  A(x)=\frac{1}{\left(  2\pi\right)  ^{\frac{3}{2}}}\int\int(e^{im_{eff}%
rch(\chi-\theta)+\vec{p}_{\perp}\vec{x}_{\perp}}a^{\ast}(p)+h.c.)\frac
{d\theta}{2}dp_{\perp}\ \\
&  \left[  a(p),a^{\ast}(p^{\prime})\right]  =2\delta(\theta-\theta^{\prime
})\delta(p_{\perp}-p_{\perp}^{\prime}),\,m_{eff}=\sqrt{\vec{p}_{\perp}%
^{2}+m^{2}}\\
&  x=(rsinh\chi,\vec{x}_{\perp},cosh\chi),\;p=(cosh\chi,\vec{p}_{\perp
},m_{eff}sinh\theta)\nonumber
\end{align}
The limit $r\rightarrow0$ together with a compensating limit $\chi=\hat{\chi
}-lnm_{eff}r$ provides a finite lightfront limit in terms of the same
creation/annihilation operators and hence takes place in the same Hilbert
space (unlike the scaling limit used for critical phenomena) and leads to the
desired result%
\begin{align}
A(x)|_{LF} &  =\frac{1}{\left(  2\pi\right)  ^{\frac{3}{2}}}\int_{0}^{\infty
}\int(e^{ip_{-}x_{+}+ip_{\perp}x_{\perp}}a^{\ast}(p)+h.c.)\frac{dp_{-}}%
{2p_{-}}dp_{\perp}\label{res}\\
p_{-} &  \simeq e^{-\theta}\nonumber
\end{align}
which yields the above formula for the two point function. The
infrared-divergence in the longitudinal factor is spurious if one views the
lightfront localization in the setting of modular wedge
localization\footnote{By re-expressing the rapidity testfunction space in
terms of the $p_{+}$ integration variable, one obtains the vanishing of the
testfunctions at $p_{+}=0.$ The same argument also shows that an additive
modification of $\hat{\chi}$ (a multiplicative change of $p_{-}$) does not
change the result in the appropriate test function setting.
\par
{\small {}}}. On the other hand the obstruction resulting from the large
$\kappa$ divergence of the K-L spectral function (short distance regime of
interaction-caused vacuum polarization) is shared with that which limits the
range of validity of the ETCR formalism. But whereas equal time restricted
interacting fields in $d\eqslantgtr1+2$ simply do not exist, there is no such
limitation on the short distance properties of generalized chiral conformal
fields which turn out to generate the ALH. What breaks down is the idea that
these lightfront generating fields can be gotten simply by restricting the
fields of the ambient theory as in the above example of free fields.

It turns out that modular theory provides a useful tool to analyze the
connection between the ambient theory and its holographic projection. Although
the ambient theory may well be given in terms of pointlike fields and the ALH
may also allow a pointlike description (see \ref{gen}), there is no known
direct relation between these fields. This was of course precisely the
unresolved problem of lightcone quantization. Even in the above
interaction-free case when the restriction method in the sense of (\ref{res})
works, the ALH net of algebras turns out to be nonlocal relative to the
ambient algebra and hence the recovery of the ambient from the ALH involves
nonlocal steps which the standard formalism cannot handle. Whereas lightcone
quantization was not able to address those subtle problems, ALH solves them.

The intuitive physical basis of this algebraic approach is a limiting form of
the causal closure property. Let $O$ be a spacetime region and $O^{\prime
\prime}$ its causal closure (the causal disjoint taken subsequently taken
twice) then the causal closure property is the following equality
\begin{equation}
\mathcal{A}(\mathcal{O})=\mathcal{A}(\mathcal{O}^{\prime\prime})
\end{equation}
In the case of free fields this abstract algebraic property\footnote{This
equality is the local version of the ``time slice property'' \cite{H-S}.} is
inherited via quantization from the Cauchy propagation in the classical
setting of hyperbolic differential equations. The lightfront is a limiting
case (characteristic surface) of a Cauchy surface. Each lightray which passes
through $O^{\prime\prime}$ either must have passed or will pass through $O.$
For the case of a $x^{0}-x^{3}$ wedge $W$ and its $x^{0}-x^{3}=0$ (upper)
causal lightfront boundary $LFB(W)$ (which covers half of a lightfront) the
relation
\begin{equation}
\mathcal{A}(LFB(W))=\mathcal{A}(W)\subset B(H)\label{LFB}%
\end{equation}
is a limiting situation of the causal shadow property; a lightlike signal
which goes through this boundary must have passed through the wedge (or in the
terminology of causality, the wedge is the backward causal completion of its
lightfront boundary). Classical data on the lightfront define a characteristic
initial value problem and the smallest region which generates data localized
in an open ambient region is half the lightfront as in (\ref{LFB}); for any
transversely not two-sided infinite extended subregion $O_{LF}$ on the
lightfront, as well as for any region on the lightfront which is bounded in
the lightray direction, the causal completion is trivial i.e. $O_{LF}%
=O_{LF}^{\prime\prime}.$ This unusual behavior of the lightfront is related
the fact that as a manifold with its metric structure inherited from the
ambient Minkowski spacetime it is not even locally hyperbolic.

Several symmetries which the lightfront inherits from the ambient Poincar\'{e}
group are obvious; it is clear that the lightlike translation together with
the two transverse translation and the transverse rotation are leaving the
lightfront invariant and that the longitudinal Lorentz boost, which leaves the
wedge invariant, acts as a dilatation on the lightray in the lightfront. There
are however two additional invariance transformations of the lightfront which
are less obvious. Their significance in the ambient space is that of the two
``translations'' in the 3-parametric Wigner little group $E(2)$ of the light
ray in the lightfront (a Euclidean subgroup of the 6-parametric Lorentz
group). Projected into the lightfront these ``translations'' look like
transverse Galilei transformations in the various $(x_{\perp})_{i}-x_{+}$ planes.

Modular concepts (in particular modular inclusions and intersections) provide
a firm operator algebraic basis for the interplay between the ambient
causality and the localization structure as well as the 7-parametric symmetry
of the lightfront holography\footnote{For the inverse holography the
information from the fundamental lightfront inclusion $A(LF(W))\subset
_{G_{LF}}A(LF)=B(H)$ has to be complemented by the action of the $x_{-}%
$translation (similar to the Hamiltonian input in the ETCR approach). }. Among
the many structural consequences we only collect those which are important for
the constructive use of holography:

\begin{itemize}
\item  The Poincar\'{e} group $P(4)$ and hence also the 7-parametric subgroup
$G_{LF}\subset P(4)$ which leaves the lightfront invariant are of modular
origin. The full lightfront symmetry is much larger and includes the Moebius
group extension of the 2-parametric longitudinal translation-dilation group
(which also turns out to be of modular origin).

\item  The lightfront algebra has no vacuum fluctuations in transverse
direction i.e. the operator algebra of an longitudinal infinitely extended
cylindrical region $\Xi=\{x_{\perp}\in Q,\,-\infty<x_{+}<\infty$\} with finite
transverse extension $Q$ is a tensor factor of the full lightfront algebra
which is identical to the full ambient algebra $B(H)$%
\begin{equation}
\mathcal{A}(LF)=B(H)=\mathcal{A}(\Xi)\otimes\mathcal{A}(\Xi)^{\prime
},\;\mathcal{A}(\Xi)^{\prime}=\mathcal{A}(\Xi^{\prime})
\end{equation}
In longitudinal direction the cyclicity of the vacuum (the Reeh-Schlieder
property) prevents such a factorization i.e. the lightfront holography
``squeezes'' the field theoretic vacuum fluctuation into the lightray
direction so that the transverse structure becomes purely quantum mechanical.
As a result of the Moebius covariance along the lightray and the quantum
mechanical factorization in transverse direction the lightfront holography has
the structure of an (quantum mechanically) extended chiral QFT.
\end{itemize}

For the derivation of the local net structure of the lightfront theory in the
longitudinal and transverse directions we refer to \cite{cross2}\cite{B-M-S};
this is the part which requires the use of modular localization concepts
(modular inclusions and modular intersections of wedge algebras, relative
commutants) which differ significantly from concepts of the standard approach
to QFT. The following remarks are only intended to facilitate understanding
and highlight some consequences.

Although there is presently no rigorous proof, the structural analogy of the
lightfront holographic projection with chiral theory leads one to expect that
similar group theoretical arguments as in \cite{Joerss} provide the existence
of covariant pointlike generators. In cases where they exist, their
commutation relations are severely restricted; the transverse quantum
mechanical nature only permits a delta function without derivative and the
balance in the scaling dimensions restricts the longitudinal singularity
structure to that of Lie fields known from chiral current or W-algebras.%

\begin{align}
&  \left[  \psi_{i}(x_{\perp},x_{+}),\psi_{j}(x_{\perp}^{\prime},x_{+}%
^{\prime})\right]  =\label{gen}\\
&  =\delta(x_{\perp}-x_{\perp}^{\prime})\left\{  \delta^{(n_{ij})}(x_{+}%
-x_{+}^{\prime})+\sum_{k}\delta^{(n_{ijk})}(x_{+}-x_{+}^{\prime})\psi
_{k}(x_{\perp},x_{+})\right\}  \nonumber
\end{align}
The classification of extended chiral theories is an open problem. As in the
pure chiral case one may hope for rational situations in which there exists a
finite set of generating fields.

The difficult part of a constructive proposal of the lightfront holography is
the ``inverse holography'' i.e. the reconstruction of an ambient theory from
its holographic projection. Apart from the interaction-free case which is
characterized by a c-number commutator, the existence and uniqueness of an
``ambient reading'' of a given extended chiral model is unknown. The analogy
with the canonical formalism suggests to expect that only with additional
``dynamical'' information e.g. the action of the $x_{-}$ lightray translation
on the lightfront net or on its generating fields (\ref{gen}) one can expect
uniqueness of the holographic inversion. 

In the remainder of this section contains some comments on the inverse
holography of factorizing models, where as a result of the two-dimensionality
the transverse structure is absent and the holographic projection is a bona
fide chiral theory. The on-shell aspect of covariant PFG generators for wedge
algebras (\ref{PFG}) trivializes the passing between the ambient theory and
its holographic projection; within the setting of factorizing models the
holographic inversion is unique and amounts to representing the action of the
$x_{-}$ translation (similar to the case of free fields) by the multiplication
with $e^{ip_{+}x_{-}},$ $p_{+}=\frac{m^{2}}{p_{-}}$ in the formfactor
representation (\ref{series}). The reason for this uniqueness is that the
covariance property of the particle-like Z-F creation/annihilation operators
implicitly fix the transformation properties of the full Poincar\'{e} group
i.e. including the LF-changing transformations beyond $G_{LF}.$

The holographic restriction of factorizing models also highlights a new
interesting aspect of chiral theories. At least those chiral models which
originate in this way permit a formal representation in terms of PFGs
(\ref{series}) inherited from ambient theory. Although this basis of
$Z^{\#}(\theta)$ operators looses its particle interpretation in the chiral
holographic projection, it still continues to provide an unexpected
``on-shell representation'' simplicity for these chiral algebras. In this
representation the Moebius rotations applied to states $Z^{\ast}(\theta
)\Omega$ ``dress'' the latter with a vacuum polarization cloud. Again we refer
for more details to \cite{cross2}\cite{B-M-S}. Analogous to the free field
case the inverse holography in this particular representation just amounts to
multiplication of the formfactors $a_{n}$ with the appropriate $e^{ip_{+}%
x_{-}}$ translation factors. The possibility that each chiral theory may have
one factorizable representative in its inverse holography equivalence class
would have drastic implications for the complete classification of chiral theories.

This invere holography also raises the interesting question about the possible
dynamical role of modular generated non-local ambient symmetries beyond the
local vacuum preserving Poincar\'{e} transformations. This is part of a quest
for a more profound future understanding of the relation between particle
aspects of the ambient theory and chiral field aspects of its holographic projection.

We conclude with some additional remarks about the difference in the
underlying philosophy as compared to the standard approach to QFT which is
based on quantizing classical field theories i.e. on the idea that important
models of particle physics can be constructed by subjecting the classical
Lagrangian formalism to quantization rules. This setting leads to a finite
number of possibilities of renormalizable local coupling between higher
dimensional ($d\geq1+2$) covariant fields which contains the important
standard gauge theory model of electro-weak interactions.

The modular based approach advocated here disposes of the parallelism to
classical field theories; instead of quantizing concrete classical field
models it aims at a classifying of models according to their intrinsic
algebraic structure. The prototype situation is that of chiral models on the
lightray which are classified by their Lie-type commutation structure or
alternatively by analyzing the possible modular position of three MA
\cite{Wies}. The underlying philosophy is that of universality classes, as it
is successfully used in the condensed matter physics of critical phenomena.
Instead of trying to find a unique model of particle physics (a ``TOE'') by
quantizing a selected classical Lagrangian, one classifies holographic
equivalence classes and refines the search for the best mathematical
description of particle physics in terms of local quantum theory by adding
additional dynamical informations.

\textit{Acknowledgement: }I have profited from several discussions with Jens
Mund who used similar modular concepts in his contribution \cite{Munds} to the
symposium in honor of Jacques Bros.

\end{document}